# Simulation of Chemical Engineering Memristive Biosensor

Manel Bouzouita
*Electronics and Microelectronics Lab*
*Faculty of Sciences Of Monastir*
*University of Monastir*
Monastir, Tunisia
Manel.Bouzouita@ugent.be

Fakhreddine Zayer
*KU Center of Autonomous Robotic Systems*
*Khalifa Univesity*
Abu Dhabi, UAE
Fakhreddine.Zayer@ku.ac.ae

Hamdi Belgacem
*Electronics and Microelectronics Lab*
*Faculty of Sciences Of Monastir*
*University of Monastir*
Monastir, Tunisia
Belgacemhamdi@gmail.com

*Abstract*—This paper introduces a perspective approach for simulating a memristive sensor tailored for low-power biological analyte detection. The necessity for such innovation stems from the increasing demand for efficient biosensing technologies that can operate with minimal power consumption. Within this study, a numerical dynamic memristive model serves as a basis platform for implementing an enhanced nano-sensing method characterized by low cost and high sensitivity. Numerous simulations were conducted to validate the suitability of the dynamic memristive model's behavior for emulating a chemical sensing approach. The simulated data is collected for deploying an AI application to ensure an advanced predictable biosensing intake function. All in all, this work paves the way for developing compact numerical models of memristive biosensors, addressing the pressing need for portable, low-power consumption biosensing solutions.

## I. INTRODUCTION

Memristors as fourth passive circuit components, often show long state lifetime and ultra-fast switching by utilizing relatively small biases, increasing switching speed by up to six orders of magnitude due to their highly non-linear switching rate [1]. The rapid switching capability is among the reasons that the ongoing attempts are determined to integrate memristors in different fields specifically the biosensing domain [2]. Despite the advancements in the biosensing field, traditional technologies continue to encounter energy and time consumption challenges that restrict their effective operation and utility in various applications [3].

In our previous work, we studied the dynamics of a simulated memristive biosensing model [4]. However, in this study, we focus on simulating memristive bio-adsorption, emulating a chemical sensor for highly sensitive, fast-responding, and cost-effective nano-detection. The simulation data is extracted for implementing an AI-based predictable adsorption model ensuring the creation of an innovative direction in the memristive biosensing field.

The body of the paper is organized as follows. Section II reviews the state of the art of memristive technologies and the important milestones. Section III reports the state of the art of the memristive biosensing concept. Section IV emphasizes the simulation prospect of chemical fluidic sensor. The results of the approach investigation are shown and discussed in Section V. Finally, section VI concludes the study and outlines our future perspectives.

## II. STATE OF THE ART

### A. Memristor

Since their fabrication within HP labs in 2008 [5], memristors have gained the interest of researchers as they overtake classical von Newman architectures in terms of rapidity [6], power [7], and density [8]. Indeed, while conventional biosensors typically exhibit significantly higher power consumption [9], [10], memristive devices operate with substantially reduced power requirements. [11]. As such, various applications have been developed, citing, memristive artificial neural networking (ANN) [12], memristive biosensing [13] and analogue computing, particularly in the context of neuromorphic computing aimed at hardware security [14]. Fig. 1 presents the progression of memristive applications within the context of the annual number of publications.

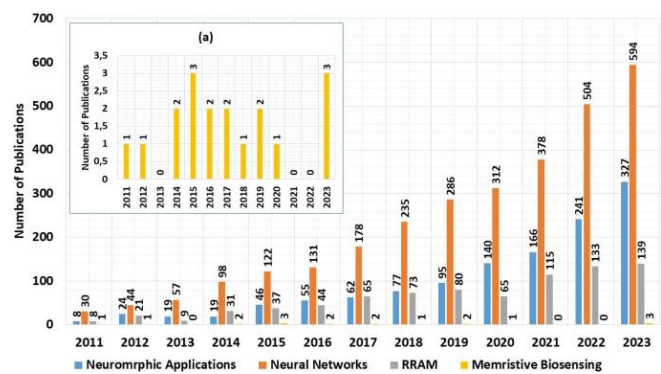

Fig. 1. Memristive applications: number of publications per year (dataset is extracted From Scopus), **(a)** Number of publications of memristive biosensing.

### B. Memristive Bio-sensing

The memristive biosensor is a nonvolatile memory nanoscaled module capable of sensing bio-molecules once functionalized [15]. One of the most common structures is made of memristive bio-functionalized Schottky barrier silicon nanowires where the resistance change involves the modulation of Schottky barriers and surface charge difference due to bio-receptor interactions [16]. In this case, the sensing mechanism detects changes in surface potential or charge distribution on silicon nanowires. Generally, the sensing process shows a voltage gap varying relatively to both functionalization and detection processes [17]. This kind is often characterized by a bipolar memristive behaviour corresponding to a couple of back-to-back Schottky diodes in contrarily functioning [18]. Fig. 2 describes the architecture of a conventional Schottky barrier silicon nanowires-based memristive biosensor [15].

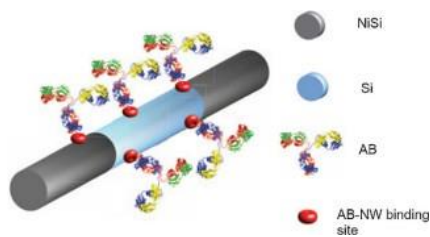

Fig. 2. Schottky barrier memristive silicon nano-wire [15].

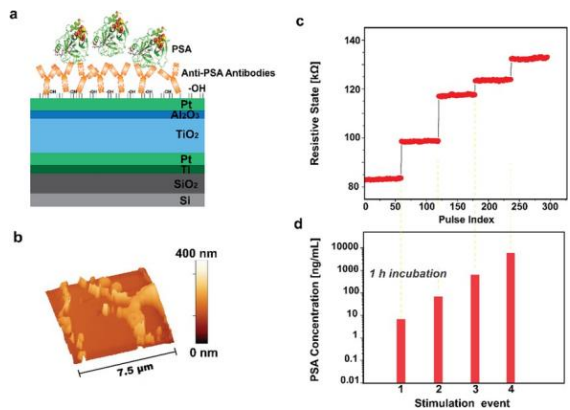

Fig. 3. Metal oxide memristive biosensor [19]. **(a-b):** model structure. **(c-d):** resistive state variation VS input bias.

Another prevalent memristive biosensing systems based on metal oxide memristors exist. They operate with the variation in resistive states during both bio-receptor fixation and bio-analyte intake [19]. Indeed, the variation of the metal oxide's resistance provokes a change in the memristor's resistive state. Furthermore, the interactions of the memristor with the fixed bio-receptors and bio-analytes, result in the modification of the resistive states of the memristor, hence, the memristive bio-sensor [20]. Typically, this memristive biosensor features a vertical structure comprising a metal oxide layer positioned between two metal electrodes, with the possible inclusion of a conductive filament [19]. Fig. 3 describes the architecture of a metal oxide-based memristive biosensor and the effect of the increase of the antigen intake [PSA] [19].

Digging deeper into the history, in 2011 David sacchetto et al. introduced memristive properties for biosensing through functionalized Schottky-barrier Silicon nanowires [13]. The study proved that the voltage-current hysteresis characterization appears after the functionalization and shrinks with the bio-molecule intake. Moreover, in 2012, Sandro Carrara and his team emphasized a new detection for dried samples using memristive silicon contacts that were fabricated with a lithography process [15]. Further theoretical analysis showed detection sensitivity is equal to 37± 1 mV/ fM and the limit of detection is around 3,4 ±1,8 fM. In 2014, the team focused on observing the humidity effect on the hysteretic voltage gap of bio-modified silicon wires [16]. In 2016, Alessandro Vallero et al. proposed the first memristive biosensing platform integrated with a microfluidic structure [21]. This study reports a fast label-free data collection with high-reliability features. In the same year, a study conducted by Ioulia Tzouvadaki claimed the elaboration of an ultra-sensitive apta-sensor for Prostate-Specific antigen (PSA) monitoring with the best limit of detection (23 aM) within apta-sensing field [22]. In 2017, a memristive crossbar-based gas sensor was introduced by Adedotun Adeyemo et al [20]. In 2018, Ioulia Tzouvadaki et al. proposed a single-chip structure for multiple bio-sensing frameworks with fast, simultaneous, and automatic output data [23]. Furthermore, in 2019, Ioulia Tzouvadaki presented a multi-panel for single or multiple markers' sensing from one sample in-closing one or numerous molecules with a remarkable limit of detection that is around 1.6 fM [24]. Equally important, in 2020, Ioulia Tzouvadaki worked on a metal oxide memristive device for PSA concentration's bio-detection with 0.6 ng/ ml as a limit of detection [19]. Last but not least, in 2023, Zrinski et al. conducted a study about a memristive crossbar array for liquid D-glucose hydroxide sensing with a detection range that varies between 10 mM and 80 mM [25]. The table " I" below represents a critical summary of the limit of detection of some memristive sensing technologies.

TABLE I
SAMPLES OF MEMRISTIVE BIO-SENSING'S LIMIT OF DETECTION

| Study | Limit of Detection | Year | Ref. |
|---|---|---|---|
| Memristive Biosensors: A New Detection Method By Using Nanofabricated Memristors | 3.4 ± 1.8 fM | 2012 | [15] |
| Memristive Biosensors Under Varying Humidity Conditions | 0.6 to 1.2 fM | 2014 | [16] |
| Label-Free Ultrasensitive Memristive Aptasensor | 23 aM | 2016 | [22] |
| Fabrication And Characterization Of Simple Structure Fluidic-Based Memristor For Immunosensing Of NS1 Protein Applications | 52nM | 2020 | [26] |
| Anodic Hfo2 Crossbar Arrays For Hydroxide-Based Memristive Sensing In Liquids: Original Scientific Paper | 10mM to 80mM | 2023 | [25] |

## III. SIMULATION OF AI-BASED MEMRISTIVE BIOSENSOR

### A. Modelling of Memristor Device

Fig. 4 illustrates a 2D and 3D dynamic metal oxide memristive models that refer to the study of Kim Sungho et al. [27]. This structure is employed for the investigation of our study goal.

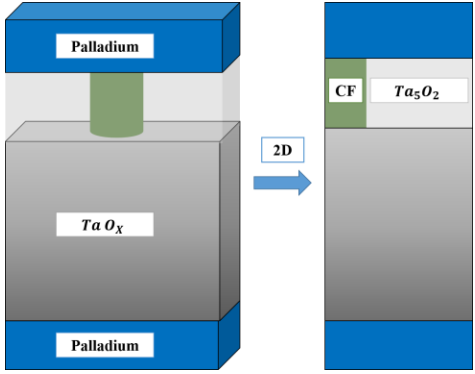

Fig. 4. Structure of the simulated metal oxide memristive device.

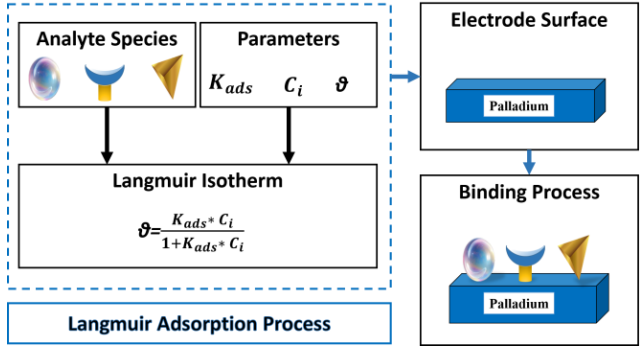

Fig. 5. Langmuir adsorption assumption for analyte fixation on the top electrode of the memristive device.

The model includes the integration of highly resistive $Ta_2O_5$ and less resistive $TaO_x$ layers positioned between two palladium electrodes, namely the word line (WL) responsible for memory signal transmission and the bit line (BL) serving as the ground reference [28], [29], [30]. This device adopts a unipolar RRAM switching behaviour of the $Pd/Ta_2O_5/TaO_x/Pd$ device which is widely acknowledged to involve the formation and rupture of the metal CF where it forms between the upper and lower Pd electrodes. In the absence of the CF, the RRAM remains in a high-resistive state (HRS), transitioning to a low-resistive state (LRS) upon CF formation. The memristor model is characterized by numerous variables and physical fundamentals which are summed in (1), (2) and (3) respectively presenting, the oxygen vacency transport, the current continuity and the joule heating equations. For the oxygen vacency transport, the following mass balance equation (1) is introduced and it implicates transport mechanisms such as diffusion and convection.

$$\frac{\delta c_i}{\delta t} + \nabla j + u \nabla c_i = R_i \quad (1)$$

In this equation, $c_i$ presents the species concentration ($mol/m^3$), $R_i$ is the reaction rate ($mol/(m^3 \cdot s)$), while u is the mass averaged velocity vector (m/s), and i is the mass diffusive flux vector ($mol/(m^2 \cdot s)$). Moreover, the electric continuity is described with the equation (2).

$$\nabla \sigma \nabla \psi = 0 \quad (2)$$

With $\sigma$ is the electric conductivity (S/m), j is the current density ($A/m^2$), and $\psi$ is the electric potential (V). For the resistive heating, an elliptic partial differential equation for the temperature is defined below:

$$\rho C_p \frac{\delta T}{\delta t} + \nabla q = Q \quad (3)$$

q refers to the conductive heat flux ($W/m^2$), $q = -k_{th} \nabla T$, $k_{th}$ is the thermal conductivity (W/(m.k)), $\rho$ is the density ($kg/m^3$) while $C_p$ is the specific heat capacity at constant pressure (J/(kg·K)) and Q is the heat source ($W/m^3$).
These equations exhibit time dependency, with an initial temperature set at 300K. Furthermore, both of electrical conductivity and thermal conductivity are defined respectively with equations (4) and (5) as it is indicated below. Certainly, the parameters of those equations are inferred and subjected to certain assumptions based on Youto Nakamura's research [31].

$$\sigma = \sigma_0 \exp(\frac{-E_{ac}}{KT}) \quad (4)$$

With $\sigma_0$ is the pre-exponential factor of the electrical conductivity (S/m), $E_{ac}$ is the activation energy for conduction (eV), K is Boltzmann constant (j/K), and T refers to the temperature (K).

$$K_{th} = K_{th0}(1 + \lambda(T - T_0)) \quad (5)$$

$K_{th0}$ is the initial thermal conductivity coefficient (W/m k). In our case, The initial temperature value, $T_0 = 300K$ and $\lambda = 0.1$.

### B. Modelling of Memristive Biosensor Technology

Lately, numerous experimental and theoretical investigations have emerged, aimed at elaborating an enhanced biosensor response [32]. Typically, the theoretical biosensing mechanisms are based on mathematical isotherms and kinetics. The most common adsorption isotherms are Langmuir and Freundlich kinetics [33]. Langmuir's model serves as a simplified representation, offering a basic approximation of monolayer adsorption with no interaction between the adsorbed species. Besides, it describes a reversible mechanism onto a homogeneous surface with a finite number of identical sites [34]. Contrarily, the Freundlich approximation allows the adsorption of multi-layers and the interaction between the adsorbed species [35]. It describes heterogeneous surfaces and non-ideal adsorption behaviour [36]. Also, it implies that adsorption increases proportionally with the concentration, suggesting a non-uniform distribution of active sites [37]. In this work, a simple Langmuir approximation is employed to describe the bulk species fixation on the memristive device's top palladium electrode according to equation (6).

$$\vartheta = \frac{(K_{ads}C_i)}{(1+(K_{ads}C_i))} \quad (6)$$

In this case, $\vartheta$ describes the surface coverage conductivity (dimensionless), $C_i$ is the concentration of the analyte in the solution (mol/$m^3$), and $K_{ads}$ is the Langmuir constant (1/(M.s)). The following equation expresses the chemical reaction representing the adsorption of the analyte onto a surface "(7)":

$$Ab \xrightleftharpoons{K\cdot ads} As \quad (7)$$

In this study, Ab presents the concentration Ci (mol/$m^3$), As is the concentration of adsorbed species (mol/$m^2$), and $K_{ads}$ is the Langmuir constant (1/(M.s)). Figure 5 reports further details of the fixation reaction. The memristive device and the adsorption kinetics simulation are elaborated within COMSOL Multiphysics and the preliminary results are shown and discussed in the coming section.

## IV. RESULTS AND DISCUSSION

### A. Simulation of Dynamic Memristive Device

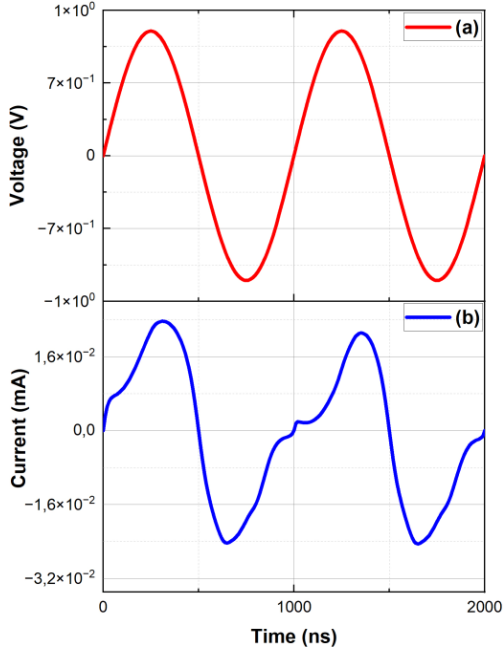

Fig. 6. The inputs and outputs of the memristor. **(a):** The applied voltage bias. **(b):** The output current variation.

The stimulation of the dynamic memristor with a sinusoidal pulse bias (figure 6 (a) ) results in distinguished preliminary outcomes. Indeed, applying a pulse train with 1.2 V as amplitude and 1000 ns as period we obtained the following figure 6 (b) that illustrates the variation of the current during a simulation time equal to 2000 ns. The observed I-V curves figure 7 display the reset and set mechanisms for bias ranging between 1V and 1.8V. While the reset transition starts at the negative side of the curve, The set begins around 1V after simulating the electro-formation of the CF as a set. The pinched hysteresis pattern intersects the origin refers to the presence of memory effect. The two hysteresis lobes are attributed to the migration of oxygen vacancies serving as charge carriers rather than phase transitions. The concentration of oxygen vacancies is modelled by solving equation (1) close to the memristor's top palladium electrode and it is crucial for the resistive switching. Specifically, the increase of the V0 concentration launches the low resistance state (LRS) imitating the formation of the CF, however, diminishing the V0 concentration leads to the decrease of the conductivity, thus, the high resistance state (HRS).

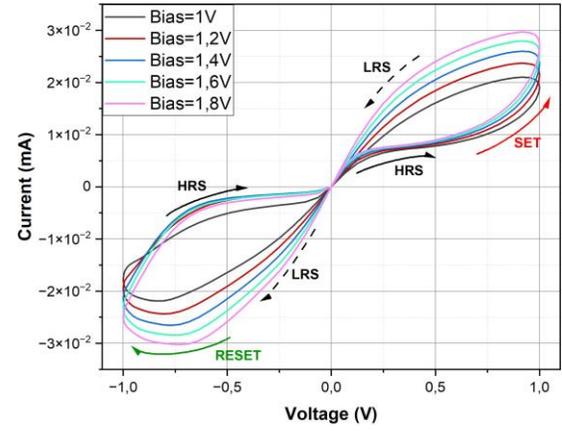

Fig. 7. The memristor's output current variation as function of different input voltage values.

### B. Simulation of Memristive Biosensor Device

The results depicted in figure 8 (blue line) illustrates the variation of the adsorbed species ($C_{ads}$) within the memristive top electrode. The increase of the ($C_{ads}$) during the simulation means the increase of the analyte intake by the sensor. The curve exhibits two distinct phases. The initial transition phase is characterised by ($C_{ads}$) ranging from 0 mol/$m^3$ to $3.10^{-7}$mol/$m^3$. The saturation phase, which initiates before 500 ns and reaches a maximum ($C_{ads}$) of approximately $3.10^{-7}$mol/$m^3$ indicates that the sensing platform is performing at a fast pace. This rapid response time observed is attributed to the swift resistive switching characteristic of the memristive technology. Furthermore, The limit of detection of this sensor is around 0.308 Umol/$m^3$.

### C. Artificial Intelligence Application for Adsorption Model Prediction

To facilitate the AI-based cleaning process, the simulated data is directly extracted from COMSOL Multiphysics utilizing the plot exporting function. A further step is deploying a machine learning (ML) model for a smart adsorption model that predicts the concentration of adsorbed species from a given time slot or array. A correlation between the raw dataset and the ML Langmuir isotherm led to a fitting fundamental parameters of the smart adsorption model according to equation (6) and Nonlinear Least Squares (NLLS) optimization. In fact, this method concludes the following values of fitting parameters where the maximum Langmuir adsorption model's capacity qmax is around $3.043 \cdot 10^{-7}$ $mol/m^3$. Furthermore, the Langmuir and time constants are respectively circling 0.0153 $m^3/mol$ and 0.657ns. Figure 8 (red stars) illustrates the extracted dataset from the simulation and the AI-predicted data array. The two curves clearly show a low root mean squared error about $6.148 \cdot 10^{-9}$ $mol^3$ , a coefficient of determination (R squared) framing 0.9615 and a high correlation between the two methods. This ML model is a perspective paradigm of a smart prototype of Langmuir-based memristive biosensing intake modelling and data fitting.

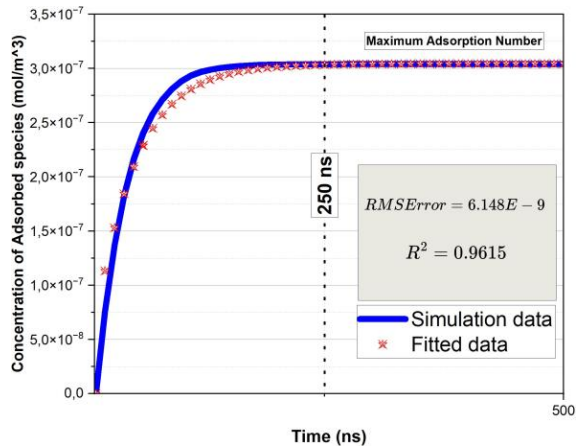

Fig. 8. Data correlation of the concentration of adsorbed species.

## V. Conclusion and Perspectives

In summary, This paper presents an innovative approach to simulating memristive sensors designed for low-power detection of biological analytes. Through the utilization of simulated memristive structures as foundational platforms, this study demonstrates the potential for implementing advanced sensing paradigms. Memristive devices emerge as promising solutions, offering a highly sensitive and cost-effective nano-sensing method. Extensive simulations validated the effectiveness of dynamic memristive models in replicating the behaviour of memristive chemical sensors. The incorporation of the sensing memristive pattern ensures the reliability of data collection and processing to facilitate AI applications.

## Acknowledgment

This work is achieved thanks to the scholar funds of the University of Monastir, the faculty of Sciences of Monastir and the Laboratory of Micro-electronics of Monastir (LR99ES30).